\documentclass[aps,twocolumn]{revtex4}
\usepackage{epsfig}
\usepackage{graphicx}
\usepackage{amsmath}
\usepackage{booktabs}
\usepackage{makecell}
\usepackage{color}

\begin{document}
\title{Color flux-tube nature of the states $T_{cs}(2900)$ and $T^a_{c\bar{s}}(2900)$}

\author{Jia Wei, Yi-Heng Wang, Chun-Sheng An${\footnote{ancs@swu.edu.cn}}$, and Cheng-Rong Deng${\footnote{crdeng@swu.edu.cn}}$}

\affiliation{School of Physical Science and Technology, Southwest University, Chongqing 400715, China}

\begin{abstract} Inspired by the states $T_{cs0}(2900)^0$, $T_{cs1}(2900)^0$,
$T^a_{c\bar{s}0}(2900)^{0}$ and $T^a_{c\bar{s}0}(2900)^{++}$ reported by the LHCb
Collaboration, we carry out a systematical investigation on the properties of the
ground and $P$-wave states $[cs][\bar{u}\bar{d}]$ and $[cu][\bar{s}\bar{d}]$ with
various spin, isospin or $U$-spin, and color combinations in a multiquark color
flux-tube model. Matching our results with the spin-parity and mass of the states $T_{cs0}(2900)^0$
and $T_{cs1}(2900)^0$, we can describe them as the compact states $[cs][\bar{u}\bar{d}]$
with $I(J^{P})=1(0^+)$ and $0(1^-)$ in the model, respectively. The ground state $T_{cs0}(2900)^0$
is mainly made of strongly overlapped an axial-vector $[cs]_{\bar{\mathbf{3}}_c}$ and
an axial-vector $[\bar{u}\bar{d}]_{\mathbf{3}_c}$. The $P$-wave state $T_{cs1}(2900)^0$
is dominantly consisted of a gradually separated scalar or axial vector $[cs]_{\bar{\mathbf{3}}_c}$
and a scalar $[\bar{u}\bar{d}]_{\mathbf{3}_c}$ in the shape of a dumbbell. Supposing the
states $T^a_{c\bar{s}0}(2900)^{0}$ and $T^a_{c\bar{s}0}(2900)^{++}$ belong to the same
isospin triplet, the mass of the state $\left [[cu]_{\bar{\mathbf{3}}_c}[\bar{s}\bar{d}]_
{\mathbf{3}_c}\right ]_{\mathbf{1}_c}$ with symmetrical $U$-spin and $J^P=0^+$ is highly
consistent with that of the states $T^a_{c\bar{s}0}(2900)^{0}$ and $T^a_{c\bar{s}0}(2900)^{++}$
in the model. After coupling two color configurations, the state $[cu][\bar{s}\bar{d}]$
is slightly lighter than the states $T^a_{c\bar{s}0}(2900)^{0}$ and $T^a_{c\bar{s}0}(2900)^{++}$.
In addition, we also discuss the properties of other states in the model.

\end{abstract}

\maketitle

\section{Introduction}

In 2020, the LHCb Collaboration observed two exotic structures
with open quark flavors in the invariant mass distribution of
$D^{-}K^+$ of the channel $B^{+}\rightarrow{D^{+}D^{-}K^{+}}$,
which were denoted as $X_{0}(2900)$ and $X_{1}(2900)$~\cite{experement}.
Their masses and widths in MeV are
\begin{eqnarray}
X_0(2900)&:& M=2866\pm7\pm2,~\Gamma=57\pm12\pm4,\nonumber\\
X_1(2900)&:& M=2904\pm5\pm1,~\Gamma=110\pm11\pm4.\nonumber
\end{eqnarray}
Both of them have the minimal quark content
$cs\bar{u}\bar{d}$ because they can strongly decay into
$D^{-}K^+$. The assignments of their spin-parity are $0^+$
and $1^-$, respectively. However, the accurate information
on their isospin has not been available until now. Recently,
the LHCb Collaboration suggested to rename the states $X_{0}(2900)$
and $X_{1}(2900)$ as $T_{cs0}(2900)^0$ and $T_{cs1}(2900)^0$,
respectively~\cite{rename}.

In 2022, the LHCb Collaboration reported two isospin vector resonances
$T^a_{c\bar{s}0}(2900)^{0}$ and $T^a_{c\bar{s}0}(2900)^{++}$ in the
$D^+_s\pi^{\pm}$ invariant spectrum of the similar channels
$B^+\rightarrow \bar{D}^-D_s^+\pi^+$ and $B^0\rightarrow\bar{D}^0D_s^+
\pi^-$~\cite{{experement1}}. Their masses and widths are
\begin{eqnarray}
T^a_{c\bar{s}}(2900)^0&:& M=2892\pm21\pm2,~\Gamma=119\pm29,\nonumber\\
T^a_{c\bar{s}}(2900)^{++}&:& M=2921\pm23\pm2,~\Gamma=137\pm35.\nonumber
\end{eqnarray}
Supposing the states belong to the same isospin triplet, the experiment also
gave the shared mass and width,
\begin{eqnarray}
 M=2908\pm23~\mbox{MeV},~\Gamma=136\pm25~\mbox{MeV}.\nonumber
\end{eqnarray}
Their least quark contents are respectively $cd\bar{s}\bar{u}$
and $cu\bar{s}\bar{d}$ with the same spin-parity $0^+$.

The investigation on the structure and property of the states could
help us to improve our knowledge of the low-energy strong
interaction. Several possible physical pictures, molecular states
$\bar{D}^*K^*$, $\bar{D}_1K$ and $D_s^*\rho$~\cite{chen2020,molina2020,liu2020,
molina2021,xiao2021,agaev2021,xue2021,liu2022,dai2022,he2021,qi2021,chenh2021,
agaev2022,rchen2022}, compact state $[cs][\bar{u}\bar{d}]$~\cite{karliner2020,
cheng2020,agaev20212,agaev20213,zhang2021,wang2021,wang2020,lv2020,ozdem2022,
yang2021,guo2022}, tetramole (superposition of molecules and compact tetraquark
states)~\cite{albuquerque2021}, triangle singularity~\cite{triangle,triangle1},
and kinematical cusp~\cite{burns}, were proposed within various theoretical
frameworks. Most of the interpretations on the states $T_{cs0}(2900)^0$
and $T_{cs1}(2900)^0$  preferred an isospin singlet. Especially for the
molecular states, the channel can produce a little of attraction by meson
exchange interaction, which is beneficial to form bound states. We refer the
interested readers to the latest reviews for more comprehensive descriptions~\cite{chen2022}.

In the present work, we prepare to make a systematical investigation
on the ground and first angular excited states $[cs][\bar{u}\bar{d}]$
and $[cu][\bar{s}\bar{d}]$ with all possible spin, isospin and color
combinations in the multiquark color flux-tube model (MCFTM). We anticipate
to broaden the property and structure of the four states from the perspective
of the diquark picture and to provide some valuable clues to the experimental
establishment of the tetraquark states in the future. We also hope that
this work can improve the understanding of the mechanism of the low-energy
strong interaction.

This paper is organized as follows. After the introduction section,
we give a concise description of the MCFTM in Sec. II. We introduce
the trial wave functions of the states $[cs][\bar{u}\bar{d}]$ in
Sec. III. We present the numerical results and discussions in Sec. IV.
We list a briefly summary in the last section.

\section{Multiquark color flux-tube model}

The multiquark color flux-tube model (MCFTM)~\cite{deng2020} has been established
on the basis of the color flux-tube picture in lattice QCD~\cite{latt1,latt2} and
the chiral constituent quark model~\cite{chiral}. We only give the schemata
of the model here. The model Hamiltonian reads
\begin{eqnarray}
\begin{aligned}
&H_n=\sum_{i=1}^n \left(m_i+\frac{\mathbf{p}_i^2}{2m_i}
\right)-T_{c}+V^{\rm CON}(n)+\sum_{i>j}^n V_{ij}, \\
&V_{ij}=V_{ij}^{\rm OGE}+V_{ij}^{\rm OBE}+V_{ij}^{\sigma},
\end{aligned}
\end{eqnarray}
where $m_i$ and $\mathbf{p}_i$ are the mass and momentum of the
$i$-th quark or antiquark, respectively. $T_c$ is the center-of-mass
kinetic energy of the states and should be deducted. $V^{\rm con}(n)$ is
an $n$-body color confinement potential. $V_{ij}^{\rm oge}$, $V_{ij}^{\rm obe}$,
and $V_{ij}^{\sigma}$ are the one-gluon-exchange interaction, the
one-boson-exchange interaction ($\pi$, $K$ and $\eta$), and the $\sigma$-meson
exchange interaction between the particles $i$ and $j$, respectively.
In the state $[cs][\bar{u}\bar{d}]$, the codes of $c$, $s$, $\bar{u}$
and $\bar{d}$ are assumed to be 1, 2, 3 and 4, respectively. Their
corresponding positions are denoted as $\mathbf{r}_1$, $\mathbf{r}_2$,
$\mathbf{r}_3$ and $\mathbf{r}_4$. The codes of the state $[cu][\bar{s}\bar{d}]$
are exactly the same as those of the state $[cs][\bar{u}\bar{d}]$.

For mesons, the quark and antiquark are linked by a three-dimensional
color flux-tube. Its two-body square confinement potential reads
\begin{eqnarray}
V^{\rm CON}(2)=kr_{q\bar{q}}^2,
\end{eqnarray}
where $r_{q\bar{q}}$ is the distance between $q$ and $\bar{q}$ and $k$
is the stiffnesses of a three-dimension color flux-tube determined
by fitting meson spectrum.

Within the framework of the diquark-antidiquark configuration, the states
$[cs][\bar{u}\bar{d}]$ and $[cu][\bar{s}\bar{d}]$ have a double-Y-type color
flux-tube structure. Its four-body confinement potential reads
\begin{eqnarray}
\begin{aligned}
V^{\rm CON}(4)=&k\left( (\mathbf{r}_1-\mathbf{y}_{12})^2
+(\mathbf{r}_2-\mathbf{y}_{12})^2+(\mathbf{r}_{3}-\mathbf{y}_{34})^2\right. \\
+&
\left.(\mathbf{r}_4-\mathbf{y}_{34})^2+\kappa_d(\mathbf{y}_{12}-\mathbf{y}_{34})^2\right),
\end{aligned}
\end{eqnarray}
where $\mathbf{y}_{12}$ and $\mathbf{y}_{34}$ stand for the positions of the two Y-shaped 
junctions. In order to satisfy the requirement of overall color singlet, the color 
flux-tube connecting $\mathbf{y}_{12}$ and $\mathbf{y}_{34}$ must be given by the SU(3) 
color representations $\bar{\mathbf{3}}_c$ or $\mathbf{6}_c$. The relative stiffness 
parameter $\kappa_{d}$ of the $d$-dimension color flux-tube is equal to 
$\frac{C_{d}}{C_3}$~\cite{kappa1,kappa2,kappa3}, where $C_d$ is the eigenvalue of the 
Casimir operator associated with the SU(3) color representation $d$ at either end of 
the color flux-tube.

Taking $\mathbf{y}_{12}$ and $\mathbf{y}_{34}$ as variational parameters,
we determine them by minimizing the four-body confinement potential. With
their values, we can obtain the minimum of the confinement potential.
Finally, we simplify the minimum into three independent harmonic oscillators
\begin{eqnarray}
V^{\rm CON}(4)=k\left(\mathbf{R}_1^2+\mathbf{R}_2^2+
\frac{\kappa_{d}}{1+\kappa_{d}}\mathbf{R}_3^2\right)
\end{eqnarray}
by diagonalizing the confinement potential matrix. $\mathbf{R}_i$ are the normal
modes of the confinement potential and read
\begin{eqnarray}
\begin{aligned}
\mathbf{R}_{1}=&
\frac{1}{\sqrt{2}}(\mathbf{r}_1-\mathbf{r}_2),~
\mathbf{R}_{2}=\frac{1}{\sqrt{2}}(\mathbf{r}_3-\mathbf{r}_4), \\
\mathbf{R}_{3}=&\frac{1}{ \sqrt{4}}(\mathbf{r}_1+\mathbf{r}_2
-\mathbf{r}_3-\mathbf{r}_4), \\
\mathbf{R}_{4}=&\frac{1}{ \sqrt{4}}(\mathbf{r}_1+\mathbf{r}_2
+\mathbf{r}_3+\mathbf{r}_4).
\end{aligned}
\end{eqnarray}

One expects the model dynamics to be governed by QCD. The perturbative
effect is the well-known one-gluon-exchange (OGE) interaction. From the
non-relativistic reduction of the OGE diagram in QCD for point-like
quarks one gets
\begin{eqnarray}
V_{ij}^{\rm OGE}={\frac{\alpha_{s}}{4}}\boldsymbol{\lambda}^c_{i}
\cdot\boldsymbol{\lambda}_{j}^c\left({\frac{1}{r_{ij}}}-
{\frac{2\pi\delta(\mathbf{r}_{ij})\boldsymbol{\sigma}_{i}\cdot
\boldsymbol{\sigma}_{j}}{3m_im_j}}\right),
\end{eqnarray}
$\boldsymbol{\lambda}^c_{i}$ and $\boldsymbol{\sigma}_{i}$ stand for the color
$SU(3)$ Gell-Mann matrices and spin $SU(2)$ Pauli matrices, respectively.
$r_{ij}$ is the distance between the particles $i$ and $j$. The Dirac
$\delta(\mathbf{r}_{ij})$ function should be regularized in the form~\cite{chiral}
\begin{equation}
\delta(\mathbf{r}_{ij})\rightarrow\frac{1}{4\pi r_{ij}r_0^2(\mu_{ij})}e^{-\frac{r_{ij}}{r_0(\mu_{ij})}},
\end{equation}
where $r_0(\mu_{ij})=\frac{\hat{r}_0}{\mu_{ij}}$, $\mu_{ij}$ is the
reduced mass of two interacting particles $i$ and $j$. The quark-gluon
coupling constant $\alpha_s$ adopts an effective scale-dependent
form given as
\begin{equation}
\alpha_s(\mu^2_{ij})=\frac{\alpha_0}{\ln\frac{\mu_{ij}^2}{\Lambda_0^2}},
\end{equation}
$\hat{r}_0$, $\Lambda_0$ and $\alpha_0$ are adjustable parameters
fixed by fitting the ground state meson spectrum.

The origin of the constituent quark mass can be traced back to
the spontaneous breaking of $SU(3)_L\otimes SU(3)_R$ chiral
symmetry~\cite{chnqm}. The chiral symmetry is spontaneously
broken in the light sector ($u$, $d$ and $s$) while it is
explicitly broken in the heavy sector ($c$ and $b$). The
meson exchange interactions only occur in the light quark
sector. The central parts of the interactions can be resumed
as follows~\cite{chiral},
\begin{eqnarray}
\begin{split}
V_{ij}^{\rm OBE}= & V^{\pi}_{ij} \sum_{k=1}^3\boldsymbol{F}_i^k
\boldsymbol{F}_j^k+V^{K}_{ij} \sum_{k=4}^7\boldsymbol{F}_i^k\boldsymbol{F}_j^k \\
+&V^{\eta}_{ij} (\boldsymbol{F}^8_i\boldsymbol{F}^8_j\cos \theta_P
-\sin \theta_P),\\
V^{\chi}_{ij}= &
\frac{g^2_{ch}}{4\pi}\frac{m^3_{\chi}}{12m_im_j}
\frac{\Lambda^{2}_{\chi}}{\Lambda^{2}_{\chi}-m_{\chi}^2}
\mathbf{\sigma}_{i}\cdot
\mathbf{\sigma}_{j}  \\
\times &\left( Y(m_\chi r_{ij})-
\frac{\Lambda^{3}_{\chi}}{m_{\chi}^3}Y(\Lambda_{\chi} r_{ij})
\right),~Y(x)=\frac{e^{-x}}{x} \\
V^{\sigma}_{ij}= &-\frac{g^2_{ch}}{4\pi}
\frac{\Lambda^{2}_{\sigma}m_{\sigma}}{\Lambda^{2}_{\sigma}-m_{\sigma}^2}
\left( Y(m_\sigma r_{ij})-
\frac{\Lambda_{\sigma}}{m_{\sigma}}Y(\Lambda_{\sigma}r_{ij})
\right).  \\
\end{split}
\end{eqnarray}
$\boldsymbol{F}_{i}$ are the flavor $SU(3)$ Gell-Mann matrices and $\chi$
represents $\pi$, $K$ and $\eta$. The mass parameters $m_{\chi}$ take
their experimental values, while the cutoff parameters $\Lambda_{\chi}$
and the mixing angles $\theta_{P}$ take the values from~\cite{chiral}.
The mass parameter $m_{\sigma}$ can be determined through the PCAC
relation $m^2_{\sigma}\approx m^2_{\pi}+4m^2_{u,d}$ ~\cite{masssigma}.
The chiral coupling constant $g_{ch}$ can be obtained from the $\pi NN$
coupling constant through
\begin{equation}
\frac{g_{ch}^2}{4\pi}=\left(\frac{3}{5}\right)^2\frac{g_{\pi NN}^2}
{4\pi}\frac{m_{u,d}^2}{m_N^2}.
\end{equation}

The most prominent characteristic is the application of the
multibody confinement potential based on the color flux-tube
picture instead of the two-body one in the other quark
models.

\section{wave functions}

Within the framework of the diquark-antidiquark configuration,
the trial wave function of the state $[cs][\bar{u}\bar{d}]$
with $I(J^P)$ can be constructed as a sum of the following direct
products of color $\varphi_c$, isospin $\varphi_i$, spin $\varphi_s$
and spatial $\phi$ terms,
\begin{equation}
\begin{split}
\Phi^{[cs][\bar{u}\bar{d}]}_{IJ}=\sum_{\alpha}c_{\alpha}
\left[\left[\left[\phi_{l_am_a}(\mathbf{r}_a)\varphi_{s_a}\right]_{J_a}
\left[\phi_{l_bm_b}(\mathbf{r}_b)\varphi_{s_b}\right]_{J_b}\right ]_{J_{ab}}\right.\\
\left.
\times\phi_{l_cm_c}(\mathbf{r}_c)\right]^{[cs][\bar{u}\bar{d}]}_{JM_J}
\left[\varphi_{i_a}\varphi_{i_b}\right]_{IM_I}^{[cs][\bar{u}\bar{d}]}
\left[\varphi_{c_a}\varphi_{c_b}\right]_{\mathbf{1}_c}^{[cs][\bar{u}\bar{d}]}~~\label{wavefunction}
\end{split}
\end{equation}
The subscripts $a$ and $b$ represent the diquark $[cs]$ and
antidiquark $[\bar{u}\bar{d}]$, respectively. The square brackets imply
all possible Clebsch-Gordan couplings. The summation index $\alpha$
represents all of possible channels and the coefficient $c_{\alpha}$
is determined by the model dynamics.

In order to obtain reliable numerical results, the precision numerical
method is indispensable. The Gaussian expansion method~\cite{GEM},
which has been proven to be rather powerful to solve few-body problem,
is therefore used in the present work. According to the Gaussian
expansion method, the relative motion wave function can be written
as
\begin{eqnarray}
\phi^G_{lm}(\mathbf{x})=\sum_{n=1}^{n_{\rm max}}c_{n}N_{nl}x^{l}
e^{-\nu_{n}x^2}Y_{lm}(\hat{\mathbf{x}}),
\end{eqnarray}
where $\mathbf{x}$ represents the Jacobian coordinates $\mathbf{r}_a$,
$\mathbf{r}_b$ and $\mathbf{r}_c$,
\begin{eqnarray}
\begin{aligned}
&\mathbf{r}_a=\mathbf{r}_1-\mathbf{r}_2,~~~\mathbf{r}_b=\mathbf{r}_3-\mathbf{r}_4, \\
&\mathbf{r}_c=\frac{m_1\mathbf{r}_1+m_2\mathbf{r}_2}{m_1+m_2}-\frac{m_3\mathbf{r}_3
+m_4\mathbf{r}_4}{m_3+m_4}
\end{aligned}
\end{eqnarray}
to describe the relative motions in the state $[cs][\bar{u}\bar{d}]$.
The corresponding angular excitations of three relative motions are,
respectively, $l_a$, $l_b$ and $l_c$. In this work, we assume that
the angular excitation only occurs between the diquark $[cs]$ and
the antidiquark $[\bar{u}\bar{d}]$ so that the $P$-parity of the
state is $(-1)^{l_c}$. The Gaussian size $\nu_n$ is taken as
geometric progression
\begin{eqnarray}
\nu_{n}=\frac{1}{r^2_n},& r_n=r_1d^{n-1},&
d=\left(\frac{r_{\rm max}}{r_1}\right)^{\frac{1}{n_{\rm max}-1}},
\end{eqnarray}
$r_1$ and $r_{\rm max}$ are the minimum and maximum of the size,
respectively. $n_{\rm max}$ is the number of the Gaussian wave
function. More details about the Gaussian expansion method
can be found in Ref.~\cite{GEM}. In the present work, we
can obtain the convergent results by taking $n_{\rm max}=7$,
$r_1=0.1$ fm and $r_{\rm max}=2.0$ fm.

The quark is in $\mathbf{3}_c$ and the antiquark is in $\bar{\mathbf{3}}_c$.
The color representation of the diquark $[cs]_{c_a}$ is antisymmetric
$\bar{\mathbf{3}}_c$ or symmetric $\mathbf{6}_c$,
\begin{center}
\begin{tabular}{c}
\end{tabular}
\begin{tabular}{|c|}
\hline
$\quad$  \\
\cline{1-1}
\hline
\end{tabular}
$\otimes$
\begin{tabular}{|c|}
\hline
$\quad$  \\
\cline{1-1}
\hline
\end{tabular}
=
\begin{tabular}{|c|}
\hline
$\quad$     \\
\cline{1-1} $\quad$   \\
\hline
\end{tabular}
$\oplus$
\begin{tabular}{|c|c|}
\hline
$\quad$ & $\quad$  \\
\cline{1-2}
\hline
\end{tabular}~.\\
\end{center}
The color representation of the antidiquark $[\bar{u}\bar{d}]_{c_b}$
is antisymmetric ${\mathbf{3}}_c$ or symmetric $\bar{\mathbf{6}}_c$,
\begin{center}
\begin{tabular}{c}
\end{tabular}
\begin{tabular}{|c|}
\hline
$\quad$  \\
\cline{1-1} $\quad$  \\
\hline
\end{tabular}
$\otimes$
\begin{tabular}{|c|}
\hline
$\quad$  \\
\cline{1-1} $\quad$  \\
\hline
\end{tabular}
=
\begin{tabular}{|c|c|}
\hline
$\quad$ & $\quad$    \\
\cline{1-2} $\quad$  \\
\cline{1-1} $\quad$  \\
\cline{1-1}
\end{tabular}
$\oplus$
\begin{tabular}{|c|c|}
\hline
$\quad$ & $\quad$  \\
\cline{1-2} $\quad$ & $\quad$ \\
\hline
\end{tabular}~.
\end{center}
According to the requirement of overall color singlet of the state
$[cs][\bar{u}\bar{d}]$, there are two ways of coupling the diquark $[cs]_{c_a}$
and antidiquark $[\bar{u}\bar{d}]_{c_b}$ into an overall color singlet:
$\left[[cs]_{\bar{\mathbf{3}}_c}[\bar{u}\bar{d}]_{\mathbf{3}_c}\right]_{\mathbf{1}}$
and $\left[[cs]_{\mathbf{6}_c}[\bar{u}\bar{d}]_{\bar{\mathbf{6}}_c}\right]_{\mathbf{1}}$,
\begin{equation}
\begin{split}
\left[[cs]_{\bar{\mathbf{3}}_c}[\bar{u}\bar{d}]_{\mathbf{3}_c}\right]_{\mathbf{1}}
=&\frac{1}{\sqrt{3}}
\left(\begin{tabular}{|c|}
\hline
            r \\
\cline{1-1} g \\
\hline
\end{tabular}
~
\begin{tabular}{|c|c|}
\hline
      r & b  \\
\cline{1-2}  g \\
\cline{1-1}  b \\
\cline{1-1}
\end{tabular}
+
\begin{tabular}{|c|}
\hline
            g \\
\cline{1-1} b \\
\hline
\end{tabular}
~
\begin{tabular}{|c|c|}
\hline
             r & r  \\
\cline{1-2}  g \\
\cline{1-1}  b \\
\cline{1-1}
\end{tabular}
+
\begin{tabular}{|c|}
\hline
            r \\
\cline{1-1} b \\
\hline
\end{tabular}
~
\begin{tabular}{|c|c|}
\hline
             r & g  \\
\cline{1-2}  g \\
\cline{1-1}  b \\
\cline{1-1}
\end{tabular}~\right),\nonumber\\
\left[[cs]_{\mathbf{6}_c}[\bar{u}\bar{d}]_{\bar{\mathbf{6}}_c}\right]_{\mathbf{1}}
=&\frac{1}{\sqrt{6}}
\left(\begin{tabular}{|c|c|}
\hline
                r   &  g      \\
\cline{1-2}
\hline
\end{tabular}
~
\begin{tabular}{|c|c|}
\hline
             r  & g      \\
\cline{1-2}  b & b  \\
\hline
\end{tabular}
+
\begin{tabular}{|c|c|}
\hline
                g   &  b      \\
\cline{1-2}
\hline
\end{tabular}
~
\begin{tabular}{|c|c|}
\hline
             r  & r      \\
\cline{1-2}  g & b  \\
\hline
\end{tabular}
+\begin{tabular}{|c|c|}
\hline
                b   &  r      \\
\cline{1-2}
\hline
\end{tabular}
~
\begin{tabular}{|c|c|}
\hline
                r  & g      \\
\cline{1-2}  g & b  \\
\hline
\end{tabular}\right.\nonumber\\
+&\left.\begin{tabular}{|c|c|}
\hline
                r   &  r      \\
\cline{1-2}
\hline
\end{tabular}
~
\begin{tabular}{|c|c|}
\hline
                g  & g      \\
\cline{1-2}  b & b  \\
\hline
\end{tabular}
+\begin{tabular}{|c|c|}
\hline
                g   &  g      \\
\cline{1-2}
\hline
\end{tabular}
~
\begin{tabular}{|c|c|}
\hline
             r  & r      \\
\cline{1-2}  b & b  \\
\hline
\end{tabular}
+\begin{tabular}{|c|c|}
\hline
   b   &  b      \\
\cline{1-2}
\hline
\end{tabular}
~
\begin{tabular}{|c|c|}
\hline
              r  & r      \\
\cline{1-2}  g & g  \\
\hline
\end{tabular}~\right).\nonumber
\end{split}
\end{equation}

The diquark $[cs]_{s_a}$ and antidiquark $[\bar{u}\bar{d}]_{s_b}$
can be in the spin singlet or triplet,
\begin{center}
\begin{tabular}{c}
\end{tabular}
\begin{tabular}{|c|}
\hline
$\quad$  \\
\cline{1-1}
\hline
\end{tabular}
$\otimes$
\begin{tabular}{|c|}
\hline
$\quad$  \\
\cline{1-1}
\hline
\end{tabular}
=
\begin{tabular}{|c|}
\hline
$\quad$     \\
\cline{1-1} $\quad$   \\
\hline
\end{tabular}
$\oplus$
\begin{tabular}{|c|c|}
\hline
$\quad$ & $\quad$  \\
\cline{1-2}
\hline
\end{tabular}~.\\
\end{center}
The total spin $S$ of the state $\left[[cs]_{s_a}[\bar{u}\bar{d}]_{s_b}\right]_S$
can be expressed as $S=s_a\oplus s_b$, its value could be 0, 1, or 2. For
the state with $S=0$, it has two coupling modes, $0=0\oplus0$ and $1\oplus1$.
Their spin wave function read
\begin{equation}
\begin{split}
\left[[cs]_{0}\oplus[\bar{u}\bar{d}]_0\right]_0&=
\begin{tabular}{|c|}
\hline
            $\uparrow$ \\
\cline{1-1} $\downarrow$ \\
\hline
\end{tabular}
~
\begin{tabular}{|c|}
\hline
             $\uparrow$  \\
\cline{1-1}  $\downarrow$ \\
\cline{1-1}
\end{tabular}~,\nonumber\\
\left[[cs]_{1}\oplus[\bar{u}\bar{d}]_1\right]_0&=
\frac{1}{\sqrt{3}}\left(\begin{tabular}{|c|c|}
\hline
            $\uparrow$ & $\uparrow$ \\
\cline{1-2}
\hline
\end{tabular}
~
\begin{tabular}{|c|c|}
\hline
      $\downarrow$ & $\downarrow$ \\
\cline{1-2}
\end{tabular}
-
\begin{tabular}{|c|c|}
\hline
            $\uparrow$ & $\uparrow$ \\
\cline{1-2}
\hline
\end{tabular}
~
\begin{tabular}{|c|c|}
\hline
      $\uparrow$ & $\downarrow$ \\
\cline{1-2}
\end{tabular}\right.\\
&+\left.
\begin{tabular}{|c|c|}
\hline
            $\downarrow$ & $\downarrow$ \\
\cline{1-2}
\hline
\end{tabular}
~
\begin{tabular}{|c|c|}
\hline
      $\uparrow$ & $\uparrow$ \\
\cline{1-2}
\end{tabular}\right),\nonumber
\end{split}
\end{equation}
where $\uparrow$ and $\downarrow$ stand for spin up and spin down,
respectively. For the state with $S=1$, it has three coupling modes,
$0\oplus1$, $1\oplus0$ and $1\oplus1$. Assuming the magnetic component
$M_s=S$, the corresponding spin wave function reads
\begin{equation}
\begin{split}
\left[[cs]_{0}\oplus[\bar{u}\bar{d}]_1\right]_1&=
\begin{tabular}{|c|}
\hline
            $\uparrow$ \\
\cline{1-1} $\downarrow$ \\
\hline
\end{tabular}
~
\begin{tabular}{|c|c|}
\hline
      $\uparrow$  & $\uparrow$ \\
\cline{1-2}
\end{tabular}~,\nonumber\\
\left[[cs]_{1}\oplus[\bar{u}\bar{d}]_0\right]_1&=
\begin{tabular}{|c|c|}
\hline
            $\uparrow$ & $\uparrow$ \\
\cline{1-2}
\hline
\end{tabular}
~
\begin{tabular}{|c|}
\hline
      $\uparrow$  \\
\cline{1-1}  $\downarrow$ \\
\cline{1-1}
\end{tabular}~,\nonumber\\
\left[[cs]_{1}\oplus[\bar{u}\bar{d}]_1\right]_1&=
\frac{1}{\sqrt{2}}\left(\begin{tabular}{|c|c|}
\hline
            $\uparrow$ & $\uparrow$ \\
\cline{1-2}
\hline
\end{tabular}
~
\begin{tabular}{|c|c|}
\hline
      $\uparrow$  &$\downarrow$ \\
\cline{1-2}
\end{tabular}
-
\begin{tabular}{|c|c|}
\hline
            $\uparrow$ & $\downarrow$ \\
\cline{1-2}
\hline
\end{tabular}
~
\begin{tabular}{|c|c|}
\hline
      $\uparrow$  &$\uparrow$ \\
\cline{1-2}
\end{tabular}\right).
\end{split}
\end{equation}
For the state with $S=2$, its spin wave function reads
\begin{equation}
\begin{split}
\left[[cs]_{1}\oplus[\bar{u}\bar{d}]_1\right]_2&=
\begin{tabular}{|c|c|}
\hline
            $\uparrow$ & $\uparrow$ \\
\cline{1-2}
\hline
\end{tabular}
~
\begin{tabular}{|c|c|}
\hline
      $\uparrow$ & $\uparrow$ \\
\cline{1-2}
\end{tabular}~.\nonumber
\end{split}
\end{equation}

The total isospin of the state is only determined by the antidiquark
$[\bar{u}\bar{d}]_{i_b}$ because of the zero isospin of the diquark
$[cs]_{i_a}$. Like the spin of the diquark or antidiquark, the
antidiquark $[\bar{u}\bar{d}]_{i_b}$ can be an isospin singlet and
triplet. The isospin wave function reads
\begin{equation}
\begin{split}
\left[[cs]_0[\bar{u}\bar{d}]_0\right]_0=cs
~\begin{tabular}{|c|}
\hline
             $\bar{u}$  \\
\cline{1-1}  $\bar{d}$ \\
\cline{1-1}
\end{tabular},~
\left[[cs]_0[\bar{u}\bar{d}]_1\right]_1=
cs~
\begin{tabular}{|c|c|}
\hline
      $\bar{u}$ &$\bar{d}$ \\
\cline{1-2}
\end{tabular}~.\nonumber
\end{split}
\end{equation}

The diquark and antidiquark are a spatially extended compound with
various color-flavor-spin-space configurations~\cite{diquark}. The
substructure of the diquarks may affect the structure of the multiquark
states. Taking all degrees of freedom of identical quarks $\bar{u}$
and $\bar{d}$ into account, the Pauli principle imposes some restrictions
on the antidiquark $[\bar{u}\bar{d}]$. $i_b+s_b=\rm even$ if the
antidiquark is in $\mathbf{3}_c$ while $i_b+s_b=\rm odd$ if the
antidiquark is in $\bar{\mathbf{6}}_c$.

The corresponding SU(2) groups of the isospin, and the so-called
$V$-spin and $U$-spin are three subgroups of the flavor SU(3) group.
The $U$-spin of the antidiquark $[\bar{s}\bar{d}]$, the $V$-spin of the
antidiquark $[\bar{s}\bar{u}]$ and the isospin of the antidiquark $[\bar{u}\bar{d}]$
have similar symmetry in their flavor wave functions. Therefore, the
total wave functions of the states $[cs][\bar{u}\bar{d}]$, $[cu][\bar{s}\bar{d}]$,
and $[cd][\bar{s}\bar{u}]$ have exactly the same structure if the
flavor SU(3) symmetry is involved. In order to avoid valueless repetition,
we just present the details of the wave function construction for the
state $[cs][\bar{u}\bar{d}]$.

\section{numerical results and analysis}

\subsection{Meson spectrum and adjustable model parameters}

We can determine the adjustable model parameters by approximately
strictly solving the two-body Schr\"{o}dinger equation to fit the ground state
meson spectrum in the MCFTM. With the Minuit program~\cite{minuit}, we can obtain
a set of optimal parameters and meson spectrum, which are presented
in Table~\ref{parameters} and~\ref{meson}, respectively.
\begin{table}[ht]
\caption{Adjustable model parameters, quark mass and $\Lambda_0$
unit in MeV, $k$ unit in MeV$\cdot$fm$^{-2}$, $r_0$ unit in MeV$\cdot$fm
and $\alpha_0$ is dimensionless.}\label{parameters}
\begin{tabular}{cccccccccccccccccc}
\toprule[0.8pt] \noalign{\smallskip}
Para.     &  ~~$m_{u,d}$~~ &  ~$m_{s}$~  &  ~~~$m_c$~~~ & ~~~$k$~~~ &  ~~$\alpha_0$~~  &~~~~~$\Lambda_0$~~~~~& ~~~$r_0$~~~  \\
\noalign{\smallskip}
Valu.     &       280      &     488     &      1653    &    458    &         3.99     &         30.34       &     65.15    \\
\toprule[0.8pt] \noalign{\smallskip}
\label{Table1}
\end{tabular}
\caption{Ground state meson spectrum, unit in MeV.\label{meson}}
\begin{tabular}{cccccccccccccccccccccccc}
\toprule[0.8pt] \noalign{\smallskip}
\renewcommand{\arraystretch}{8.0}
State     &~~~$\pi$~~~&~~$\rho$~~ &~~$\omega$~~&  ~~$K$~~ & ~$K^*$~ &~~~$\phi$~~~&~~~$D^{\pm}$   \\
\toprule[0.8pt] \noalign{\smallskip}
Theo.     &    154    &    799    &    700     &   467    &   932   &   1047   & ~~~1871    \\
PDG.      &    139    &    775    &    783     &   496    &   896   &   1020   & ~~~1869    \\
\toprule[0.8pt] \noalign{\smallskip}
States    &~~~$D^*$~~~&~~~$D_s^{\pm}$~~~&~~~$D_s^*$~~~&~~~$\eta_c$~~~&~~~$J/\Psi$~~~\\
\noalign{\smallskip}
\toprule[0.8pt] \noalign{\smallskip}
Theo.     &    2026   &    1975   &    2146    &    2977  &   3155    \\
PDG.      &    2007   &    1968   &    2112    &    2980  &   3097    \\
\toprule[0.8pt]  \noalign{\smallskip}
\end{tabular}
\end{table}

\begin{table*}[ht]
\caption{Mass of the state $[cs][\bar{u}\bar{d}]$ and contribution from each part of the Hamiltonian
unit in MeV, and the average distances unit in fm, $J=l_c\oplus S$. $\bar{\mathbf{3}}_c$-$\mathbf{3}_c$
and $\mathbf{6}_c$-$\bar{\mathbf{6}}_c$ stand for the states $[cs][\bar{u}\bar{d}]$ with the color
configurations $\left[[cs]_{\bar{\mathbf{3}}_c}[\bar{u}\bar{d}]_{\mathbf{3}_c}\right]_{\mathbf{1}}$ and
$\left[[cs]_{\mathbf{6}_c}[\bar{u}\bar{d}]_{\bar{\mathbf{6}}_c}\right]_{\mathbf{1}}$, respectively.
C.C. represents the coupling of the two color configurations. $E_k$, $V^{\rm CON}$,
$V^{\rm CM}$, $V^{\rm C}$, $V^{\eta}$, $V^{\pi}$, $V^K$ and $V^{\sigma}$ represent
kinetic energy, confinement potential, color-magnetic interaction, Coulomb interaction,
$\eta$ exchange interaction, $\pi$ exchange interaction, $K$ exchange interaction and $\sigma$ exchange interaction,
respectively.} \label{csud}
\begin{tabular}{cccccccccccccccccccccccccccccccccccccccccc}
\toprule[0.8pt] \noalign{\smallskip}
$l_c$&$S$&$IJ^{P}$&Color&Mass,~Ratio&$\langle E_k \rangle$&$\langle V^{\rm CON}\rangle$& $\langle V^{\rm CM}\rangle$&
$\langle V^{\rm C}\rangle$&$\langle V^{\eta}\rangle$&$\langle V^{\pi}\rangle$&$\langle V^{K}\rangle$&$\langle V^{\sigma}\rangle$& $\langle\mathbf{r}_{12}^2\rangle^{\frac{1}{2}}$&$\langle\mathbf{r}_{34}^2\rangle^{\frac{1}{2}}$&$\langle\mathbf{r}_{13}^2\rangle^{\frac{1}{2}}$&
$\langle\mathbf{r}_{24}^2\rangle^{\frac{1}{2}}$&$\langle\mathbf{r}_c^2\rangle^{\frac{1}{2}}$\\
\noalign{\smallskip}
\toprule[0.8pt] \noalign{\smallskip}
  &   &                        & $\bar{\mathbf{3}}_c$-$\mathbf{3}_c$ & 2559,~84\%   & 1491 & 275 & $-284$ & $-1201$ & $72$ & $-401$ &0& $-94$ &0.63 & 0.66 & 0.71 & 0.85 & 0.61\\
  & 0 & $00^{+}$               & $\mathbf{6}_c$-$\bar{\mathbf{6}}_c$ & 2830,~16\%   & 1099 & 340 & $-211$ & $-1066$ & $2$  & $38$   &0& $-73$ &0.67 & 0.89 & 0.68 & 0.85 & 0.50\\
  &   &                        & C.C.                                & 2495         & 1571 & 255 & $-409$ & $-1248$ & $63$ & $-339$ &0& $-99$ &0.60 & 0.67 & 0.66 & 0.80 & 0.56\\
\noalign{\smallskip}
  &   &                        & $\bar{\mathbf{3}}_c$-$\mathbf{3}_c$ & 2604,~98\%   & 1430 & 290 & $-235$ & $-1164$ & $72$ & $-399$ &0& $-91$ &0.67 & 0.66 & 0.72 & 0.87 & 0.62\\
0 & 1 & $01^{+}$               & $\mathbf{6}_c$-$\bar{\mathbf{6}}_c$ & 3008,~2\%    & 907  & 402 & $-5$   & $-962$  & $-3$ & $30$   &0& $-62$ &0.74 & 0.95 & 0.75 & 0.93 & 0.55\\
  &   &                        & C.C.                                & 2591         & 1444 & 285 & $-255$ & $-1173$ & $70$ & $-389$ &0& $-92$ &0.66 & 0.66 & 0.71 & 0.86 & 0.61\\
\noalign{\smallskip}
  & 2 & $02^{+}$               & $\mathbf{6}_c$-$\bar{\mathbf{6}}_c$ & 3068,~100\%  & 852  & 423 & $55$   & $-926$  & $-5$ & $27$   &0& $-59$ &0.76 & 0.97 & 0.77 & 0.95 & 0.57\\
\noalign{\smallskip}
\toprule[0.8pt] \noalign{\smallskip}
  &   &                        & $\bar{\mathbf{3}}_c$-$\mathbf{3}_c$ & 2940,~78\%   & 955  & 366 & $-12$  & $-984$  & $-2$ & $-16$  &0& $-68$ &0.67 & 0.88 & 0.78 & 0.92 & 0.62\\
  & 0 & $10^{+}$               & $\mathbf{6}_c$-$\bar{\mathbf{6}}_c$ & 3068,~22\%   & 860  & 430 & $42$   & $-935$  & $5$  & $22$   &0& $-57$ &0.75 & 1.00 & 0.78 & 0.95 & 0.56\\
  &   &                        & C.C.                                & 2871         & 1036 & 346 & $-107$ & $-1028$ & $1$  & $-6$   &0& $-72$ &0.65 & 0.87 & 0.73 & 0.88 & 0.59\\
\noalign{\smallskip}
  &   &                        & $\bar{\mathbf{3}}_c$-$\mathbf{3}_c$ & 2949,~84\%   & 955  & 370 & $1$    & $-989$  & $-6$ & $-16$  &0& $-67$ &0.64 & 0.89 & 0.80 & 0.93 & 0.64\\
0 & 1 & $11^{+}$               & $\mathbf{6}_c$-$\bar{\mathbf{6}}_c$ & 3056,~16\%   & 870  & 424 & $31$   & $-939$  & $5$  & $22$   &0& $-58$ &0.74 & 1.00 & 0.77 & 0.94 & 0.59\\
  &   &                        & C.C.                                & 2979         & 989  & 362 & $-44$  & $-1007$ & $-4$ & $-9$   &0& $-9$  &0.64 & 0.89 & 0.77 & 0.90 & 0.61\\
\noalign{\smallskip}
  & 2 & $12^{+}$               & $\bar{\mathbf{3}}_c$-$\mathbf{3}_c$ & 3018,~100\%  & 878  & 395 & $72$   & $-942$  & $-8$ & $-15$  &0& $-63$ &0.69 & 0.90 & 0.82 & 0.96 & 0.66\\
\noalign{\smallskip}
\toprule[0.8pt]
\noalign{\smallskip}
  &   &                        & $\bar{\mathbf{3}}_c$-$\mathbf{3}_c$ & 2901,~98\%   & 1468 & 391 & $-267$ & $-1014$ & $67$ & $-375$ &0& $-70$ &0.67 & 0.96 & 1.08 & 0.97 & 0.89\\
  & 0 & $01^{-}$               & $\mathbf{6}_c$-$\bar{\mathbf{6}}_c$ & 3341,~2\%    & 1007 & 502 & $-108$ & $-739$  & $1$  & $24$   &0& $-47$ &0.77 & 1.01 & 0.92 & 1.09 & 0.75\\
  &   &                        & C.C.                                & 2893         & 1478 & 385 & $-283$ & $-1017$ & $66$ & $-367$ &0& $-70$ &0.66 & 0.69 & 0.95 & 1.07 & 0.88\\
\noalign{\smallskip}
  &   &                        & $\bar{\mathbf{3}}_c$-$\mathbf{3}_c$ &  2940,~$99\%$& 1421 & 406 & $-224$ & $-987$ & $67$ & $-375$ &0& $-69$ &0.71 & 0.69 & 0.97 & 1.10 & 0.89\\
1 & 1 & $00^{-},01^{-},02^{-}$ & $\mathbf{6}_c$-$\bar{\mathbf{6}}_c$ &  3433,~$1\%$ & 912  & 545 & $-5$   & $-697$ & $-2$ & $22$   &0& $-43$ &0.81 & 1.04 & 0.97 & 1.14 & 0.80\\
  &   &                        & C.C.                                &  2938        & 1424 & 404 & $-228$ & $-987$ & $66$ & $-373$ &0& $-69$ &0.71 & 0.69 & 0.97 & 1.10 & 0.90\\
\noalign{\smallskip}
  & 2 & $01^{-},02^{-},03^{-}$ & $\mathbf{6}_c$-$\bar{\mathbf{6}}_c$ &  3464,~100\% & 882  & 557 & $26$   & $-679$ & $-3$ & $21$   &0& $-41$ &0.81 & 1.05 & 0.99 & 1.16 & 0.81\\
\noalign{\smallskip}
\toprule[0.8pt]
\noalign{\smallskip}
  &   &                        & $\bar{\mathbf{3}}_c$-$\mathbf{3}_c$ & 3275,~94\%   & 967  & 491 & $13$   & $-831$  & $-4$ & $-14$  &0& $-48$ &0.71 & 0.93 & 1.01 & 1.14 & 0.87\\
  & 0 & $11^{-}$               & $\mathbf{6}_c$-$\bar{\mathbf{6}}_c$ & 3475,~6\%    & 886  & 566 & $33$   & $-690$  & $3$  & $16$   &0& $-40$ &0.82 & 1.09 & 0.99 & 1.16 & 0.80\\
  &   &                        & C.C.                                & 3259         & 990  & 582 & $-16$  & $-834$  & $-3$ & $-12$  &0& $-49$ &0.70 & 0.93 & 0.99 & 1.12 & 0.86\\
\noalign{\smallskip}
  &   &                        & $\bar{\mathbf{3}}_c$-$\mathbf{3}_c$ & 3265,~98\%   & 987  & 489 & $1$    & $-846$  & $-5$ & $-14$  &0& $-48$ &0.67 & 0.93 & 1.02 & 1.14 & 0.90\\
1 & 1 & $10^{-},11^{-},12^{-}$ & $\mathbf{6}_c$-$\bar{\mathbf{6}}_c$ & 3471,~2\%    & 892  & 565 & $25$   & $-691$  & $3$  & $16$   &0& $-40$ &0.80 & 1.08 & 0.98 & 1.15 & 0.80\\
  &   &                        & C.C.                                & 3259         & 994  & 487 & $-9$   & $-847$  & $-5$ & $-13$  &0& $-48$ &0.67 & 0.93 & 1.02 & 1.13 & 0.90\\
\noalign{\smallskip}
  & 2 & $11^{-},12^{-},13^{-}$ & $\bar{\mathbf{3}}_c$-$\mathbf{3}_c$ & 3316,~100\%  & 929  & 510 & $56$   & $-814$  & $-6$ & $-14$  &0& $-46$ &0.71 & 0.93 & 1.04 & 1.17 & 0.92\\                                                                                          \noalign{\smallskip}
\toprule[0.8pt]
\end{tabular}
\end{table*}

\subsection{$[cs][\bar{u}\bar{d}]$ spectrum and candidates of $T_{cs}(2900)$}

In the following, we concentrate on the properties of the ground and $P$-wave states
$[cs][\bar{u}\bar{d}]$ with various spin, isospin and color combinations in the MCFTM
with the parameters determined by the meson spectrum. Note that we do not introduce any
new adjustable parameters in the calculation of the tetraquark states.

Solving the four-body Schr\"{o}dinger equation with the well-defined trial wave function,
we can obtain the eigen energies of the states $[cs][\bar{u}\bar{d}]$, which are presented
in Table~\ref{csud}. $\bar{\mathbf{3}}_c$-$\mathbf{3}_c$ and $\mathbf{6}_c$-$\bar{\mathbf{6}}_c$
respectively stand for the state $[cs][\bar{u}\bar{d}]$ with the color configurations
$\left[[cs]_{\bar{\mathbf{3}}_c}[\bar{u}\bar{d}]_{\mathbf{3}_c}\right]_{\mathbf{1}}$
and $\left[[cs]_{\mathbf{6}_c}[\bar{u}\bar{d}]_{\bar{\mathbf{6}}_c}\right]_{\mathbf{1}}$.
C.C. represents the coupling of the two color configurations. We calculate the ratio of
each color configuration in the coupled channels using the corresponding eigenvectors.
In addition, we calculate the contribution coming from each part of the Hamiltonian in
each color configuration and the coupled channels, which are presented in each row in Table~\ref{csud}.

In order to illustrate the spatial configuration of the states, we also calculate
the average distance between two quarks $\langle\mathbf{r}_{ij}^2\rangle^{\frac{1}{2}}$
and the relative distance $\langle\mathbf{r}_c^2\rangle^{\frac{1}{2}}$ between
the diquark $[cs]$ and the antidiquark $[\bar{u}\bar{d}]$, which are listed in
Table~\ref{csud}. $\langle\mathbf{r}_{12}^2\rangle^{\frac{1}{2}}$ and
$\langle\mathbf{r}_{34}^2\rangle^{\frac{1}{2}}$ represent the size of the
diquark $[cs]$ and antidiquark $[\bar{u}\bar{d}]$, respectively.
$\langle\mathbf{r}_{13}^2\rangle^{\frac{1}{2}}$ is equal to
$\langle\mathbf{r}_{14}^2\rangle^{\frac{1}{2}}$
and $\langle\mathbf{r}_{23}^2\rangle^{\frac{1}{2}}$ is equal to
$\langle\mathbf{r}_{24}^2\rangle^{\frac{1}{2}}$
because the quarks $\bar{u}$ and $\bar{d}$ are considered as identical particles. All of
the distances are less than or around 1 fm so that the states $[cs][\bar{u}\bar{d}]$
should be a compact spatial configuration in the model because of the multi-body
confinement potential, which is a collective degree of freedom and bind all
particles together.

For the ground states, the diquark $[cs]$ and the antidiquark $[\bar{u}\bar{d}]$ have
a strongly overlap because of the smaller distance $\langle\mathbf{r}_c^2\rangle^{\frac{1}{2}}$
relative to the sizes of the diquark $[cs]$ and antidiquark $[\bar{u}\bar{d}]$, see
$\langle\mathbf{r}_{12}^2\rangle^{\frac{1}{2}}$,
$\langle\mathbf{r}_{34}^2\rangle^{\frac{1}{2}}$,
and $\langle\mathbf{r}_{c}^2\rangle^{\frac{1}{2}}$. For the $P$-wave states,
the sizes of the diquark $[cs]$ and antidiquark $[\bar{u}\bar{d}]$ do not change
dramatically relative to those of the corresponding ground states because the
angular excitation only occurs between the diquark $[cs]$ and antidiquark
$[\bar{u}\bar{d}]$. However, the distance between the diquark $[cs]$ and
antidiquark $[\bar{u}\bar{d}]$ obviously increase, also see
$\langle\mathbf{r}_{12}^2\rangle^{\frac{1}{2}}$, $\langle\mathbf{r}_{34}^2\rangle^{\frac{1}{2}}$,
and $\langle\mathbf{r}_{c}^2\rangle^{\frac{1}{2}}$. The $P$-wave states look
like a dumbbell-like spatial configuration because the $[cs]$ and $[\bar{u}\bar{d}]$
is separated gradually.

One can find from Table~\ref{csud} that the $\bar{\mathbf{3}}_c$-$\mathbf{3}_c$
is dominant in the states with $S=0$ and 1, especially for the states $[cs][\bar{u}\bar{d}]$
with $I=0$. In the $\bar{\mathbf{3}}_c$-$\mathbf{3}_c$, the interactions
$V^{\rm cm}$, $V^{\rm C}$ and $V^{\pi}$ can give much stronger attractions
than they do in the $\mathbf{6}_c$-$\bar{\mathbf{6}}_c$. With the increasing
of the mass ratio of $m_Q$ and $m_{\bar{q}}$, where $Q=s$, $c$ or $b$ and
$\bar{q}=\bar{u}$ or $\bar{d}$, the $\bar{\mathbf{3}}_c$-$\mathbf{3}_c$ gradually
increase in the states $[QQ][\bar{u}\bar{d}]$~\cite{deng2020x}. The underlying
dynamical mechanism of such phenomenological regularity in the $\bar{\mathbf{3}}_c$-$\mathbf{3}_c$
is governed by the color Coulomb interaction in the diquark $[QQ]$ and
the color-magnetic interaction and the $\pi$-meson exchange in the
antidiquark $[\bar{u}\bar{d}]$~\cite{diquark}. The single color configuration
of the high-spin $(S=2)$ states is uniquely determined by the symmetry
of their wave functions.

The ground state $[cs][\bar{u}\bar{d}]$ with $I(J^P)=0(0^+)$ and
$\bar{\mathbf{3}}_c$-$\mathbf{3}_c$ has a low mass of 2559 MeV due to the strong
$\pi$-meson exchange. After coupling with the $\mathbf{6}_c$-$\bar{\mathbf{6}}_c$,
the mass of the state with $0(0^+)$ is further pushed down to 2495 MeV, which is
much lower, about 370 MeV, than that of the state $T_{cs0}(2900)^0$ reported
by the LHCb Collaboration. Therefore, the state $T_{cs0}(2900)^0$ cannot
be seen as the state $[cs][\bar{u}\bar{d}]$ with $0(0^+)$ in the model.
A similar model study on the state $[cs][\bar{u}\bar{d}]$ was carried out
in Refs.~\cite{wang2021,lv2020}, where the authors did not take into account
the meson exchange in their models. None of their predicted masses of the
state with $0(0^+)$ can match with that of the state $T_{cs0}(2900)^0$.
In other words, the state $T_{cs0}(2900)^0$ may be not the compact state
$[cs][\bar{u}\bar{d}]$ with $0(0^+)$ in the quark models with QCD-inspired
dynamics. However, various color-magnetic models without explicit dynamics
can interpret the main component of the state $T_{cs0}(2900)^0$ as the compact
state $[cs][\bar{u}\bar{d}]$ with $0(0^+)$~\cite{karliner2020,cheng2020}.
On the other hand, the color-magnetic models do not seem to completely absorb
the dynamic effect by the effective masses of the constituent quarks~\cite{deng2021}.

The ground state $[cs][\bar{u}\bar{d}]$ with $I(J^P)=0(0^+)$ is about 100
MeV higher than that of the state with $0(0^+)$ mainly due to the relative weak
color-magnetic interaction and Coulomb interaction. The $\mathbf{6}_c$-$\bar{\mathbf{6}}_c$
has a very tiny probability, just 2\%, so that it can be abandoned in the
state $[cs][\bar{u}\bar{d}]$ with $0(1^+)$. The state $[cs][\bar{u}\bar{d}]$
with $0(2^+)$ has a very high energy of 3068 MeV because of the absence of the
$\bar{\mathbf{3}}_c$-$\mathbf{3}_c$.

For the ground states with $I(J^P)=1(0^+)$ and $1(1^+)$, their masses are much
higher than the states with $0(0^+)$ and $0(1^+)$, respectively. Such
regularity also holds true for their corresponding $P$-wave states with
$I=0$ and 1, see Table~\ref{csud}, which mainly originates from their
different contribution of the $\pi$-meson exchange. This provides very
strong attraction in the states with $I=0$ while it gives a weak interaction
in the states with $I=1$. For the high-spin ($S=2$) ground states, the
mass splitting between the states with $I=0$ and $I=1$ resulting from the
$\pi$-meson exchange is not as obvious as the low-spin states.

In the ground state $[cs][\bar{u}\bar{d}]$ with $I(J^P)=1(0^+)$, its main
color configuration is $\bar{\mathbf{3}}_c$-$\mathbf{3}_c$, reaching 78\%,
and its corresponding spin configuration is $1\oplus1$, namely consisting
of an axial-vector $[cs]_{\bar{\mathbf{3}}_c}$ and an axial-vector
$[\bar{u}\bar{d}]_{\mathbf{3}_c}$, see Table~\ref{csud}. Its mass, about
2923 MeV, is slightly higher than that of the state $T_{cs0}(2900)^0$.
Taking the coupling with the $\mathbf{6}_c$-$\bar{\mathbf{6}}_c$
into account, the mass can be pushed down to 2871 MeV, which is highly
consistent with that of the state $T_{cs0}(2900)^0$. In this way, we can
describe the state $T_{cs0}(2900)^0$ as the ground state $[cs][\bar{u}\bar{d}]$
with $1(0^+)$ in the MCFTM, which is supported by the conclusions of the
similar model research and QCD sum rule~\cite{wang2021,wang2020}. If the
state $T_{cs0}(2900)^0$ really belongs to an isotriplet, its charged
partners would be abundant, which deserves further research in the future.

On the contrary, the diquark picture $[cs][\bar{u}\bar{d}]$ seems to
prefer the $I(J^P)$ assignment of $0(0^+)$ in the color-magnetic models
and QCD sum rule~\cite{karliner2020,cheng2020,zhang2021}. Assuming the
state $T_{cs0}(2900)^0$ is determined to be isosinglet eventually, the
molecular configuration $\bar{D}^*K^*$ may be a suitable candidate
in the models. In order to discriminate all possible interpretations,
Burns \emph{et al} carried out an exhaustive analysis on their decay
behaviors as well as their productions in $B^0$ and $B^+$ decays~\cite{burns}.

In the $P$-wave states, we do not consider the spin-orbit
interaction in the present work because its contributions are very
small, just about several MeV~\cite{deng2020}. It does not change
the qualitative conclusions for the compact tetraquark states. The
spin singlet with $0(1^-)$ has a mass of 2893 MeV in the MCFTM, see
Table~\ref{csud}, which is in good agreement with that of the state
$T_{cs1}(2900)^0$. Its dominant component is composed of a scalar
$[cs]_{\bar{\mathbf{3}}_c}$ and a scalar $[\bar{u}\bar{d}]_{\mathbf{3}_c}$.
In addition, the spin triplet with $0(1^-)$ has a mass of about 2938
MeV and it consists of a scalar $[\bar{u}\bar{d}]_{\mathbf{3}_c}$ and
an axial vector $[cs]_{\bar{\mathbf{3}}_c}$. The state is not far away
from the state $T_{cs1}(2900)^0$ so that we cannot rule out the fact
that its main component may be made of a scalar
$[\bar{u}\bar{d}]_{\mathbf{3}_c}$ and an axial vector $[cs]_{\bar{\mathbf{3}}_c}$.
In other words, we can describe the state $T_{cs1}(2900)^0$ as the compact
state $[\bar{u}\bar{d}]_{\mathbf{3}_c}$ with $0(1^-)$ in the MCFTM.
Its main component could be consisted of a scalar or an axial vector
$[cs]_{\bar{\mathbf{3}}_c}$ and a scalar $[\bar{u}\bar{d}]_{\mathbf{3}_c}$.
Whichever description in the compact state $[cs][\bar{u}\bar{d}]$ and
molecular state $D_1K$, the state $T_{cs1}(2900)^0$ seems to prefer
the $I(J^P)$ assignment of $0(1^-)$~\cite{agaev20212,ozdem2022,he2021,qi2021,chenh2021}.

The states with $0(1^-)$ and $S=2$ are much higher, about 500 MeV, than
the state $T_{cs1}(2900)^0$, which should not be the main component of
the state $T_{cs1}(2900)^0$. All of the $P$-wave states with $I=1$ have
similar masses, around 3300 MeV, which are also much higher than the
state $T_{cs1}(2900)^0$. Therefore, in the MCFTM the state $T_{cs1}(2900)^0$ should
not be an isospin triplet if it is a compact state $[cs][\bar{u}\bar{d}]$.

\begin{table*}[ht]
\caption{Mass of the state $[cu][\bar{s}\bar{d}]$ and contribution from each
part of the Hamiltonian unit in MeV, and the average distances unit in fm. $U$ represents
the $U$-spin, $U=0$ and $1$ denote the antisymmetrical and symmetrical $[\bar{s}\bar{d}]$,
respectively. Other symbols have the same meanings with those in Table \ref{csud}.} \label{cusd}
\begin{tabular}{cccccccccccccccccccccccccccccccc}
\toprule[0.8pt] \noalign{\smallskip}
$l_c$&$S$&$UJ^{P}$&Color&Mass,Ratio&$\langle E_k \rangle$&$\langle V^{\rm CON}\rangle$& $\langle V^{\rm CM}\rangle$&
$\langle V^{\rm C}\rangle$&$\langle V^{\eta}\rangle$&$\langle V^{\pi}\rangle$&$\langle V^{K}\rangle$&$\langle V^{\sigma}\rangle$& $\langle\mathbf{r}_{12}^2\rangle^{\frac{1}{2}}$&$\langle\mathbf{r}_{34}^2\rangle^{\frac{1}{2}}$&$\langle\mathbf{r}_{13}^2\rangle^{\frac{1}{2}}$&
$\langle\mathbf{r}_{24}^2\rangle^{\frac{1}{2}}$&$\langle\mathbf{r}_c^2\rangle^{\frac{1}{2}}$\\
\noalign{\smallskip}
\toprule[0.8pt] \noalign{\smallskip}
  & 0 & $00^{+}$               & $\bar{\mathbf{3}}_c$-$\mathbf{3}_c$ & 2710,~60\%  & 1309 & 300 & $-206$ & $-1132$   & $-27$ & $0$   & $-148$ & $-87$ &0.73&0.66&0.64&0.95&0.59\\
  &   &                        & $\mathbf{6}_c$-$\bar{\mathbf{6}}_c$ & 2778,~40\%  & 1162 & 318 & $-236$ & $-1087$   & $5$   & $-23$ & $13$   & $-75$ &0.73&0.81&0.55&0.93&0.45\\
  &   &                        & C.C.                                & 2583        & 1461 & 262 & $-422$ & $-1213$   & $-14$ & $-12$ & $-86$  & $-94$ &0.67&0.68&0.55&0.86&0.48\\
\noalign{\smallskip}
  &   &                        & $\bar{\mathbf{3}}_c$-$\mathbf{3}_c$ & 2757,~94\%  & 1254 & 315 & $-155$ & $-1100$   & $-27$ & $0$   & $-147$ & $-84$ &0.76&0.66&0.65&0.98&0.59\\
0 & 1 & $01^{+}$               & $\mathbf{6}_c$-$\bar{\mathbf{6}}_c$ & 3003,~6\%   & 905  & 396 & $1$    & $-951$    & $2$   & $0$   & $9$    & $-60$ &0.83&0.88&0.62&1.04&0.52\\
  &   &                        & C.C.                                & 2737        & 1273 & 308 & $-186$ & $-1112$   & $-25$ & $0$   & $-137$ & $-85$ &0.75&0.67&0.63&0.96&0.57\\
\noalign{\smallskip}
  & 2 & $02^{+}$               & $\mathbf{6}_c$-$\bar{\mathbf{6}}_c$ & 3073,~100\% & 838  & 422 & $61$   & $-909$    & $1$   & $7$   & $8$    & $-56$ &0.85&0.91&0.65&1.07&0.54\\
\noalign{\smallskip}
\toprule[0.8pt]
\noalign{\smallskip}
  &   &                        & $\bar{\mathbf{3}}_c$-$\mathbf{3}_c$ & 2923,~75\%  & 978  & 354 & $-27$  & $-993$    & $5$   & $-18$ & $-18$  & $-69$ &0.74&0.81&0.66&1.00&0.58\\
  & 0 & $10^{+}$               & $\mathbf{6}_c$-$\bar{\mathbf{6}}_c$ & 3048,~25\%  & 865  & 417 & $36$   & $-929$    & $-3$  & $0$   & $17$   & $-56$ &0.83&0.92&0.63&1.06&0.52\\
  &   &                        & C.C.                                & 2837        & 1075 & 331 & $-137$ & $-1041$   & $3$   & $-15$ & $-7$   & $-73$ &0.72&0.80&0.61&0.96&0.52\\
\noalign{\smallskip}
  &   &                        & $\bar{\mathbf{3}}_c$-$\mathbf{3}_c$ & 2944,~77\%  & 942  & 368 & $-10$  & $-978$    & $3$   & $0$   & $-16$  & $-66$ &0.74&0.83&0.69&1.02&0.61\\
0 & 1 & $11^{+}$               & $\mathbf{6}_c$-$\bar{\mathbf{6}}_c$ & 3033,~23\%  & 879  & 410 & $21$   & $-936$    & $-3$  & $0$   & $18$   & $-57$ &0.82&0.92&0.63&1.05&0.52\\
  &   &                        & C.C.                                & 2907        & 987  & 357 & $-64$  & $-999$    & $1$   & $0$   & $-8$   & $-68$ &0.73&0.83&0.65&1.00&0.57\\
\noalign{\smallskip}
  & 2 & $12^{+}$               & $\bar{\mathbf{3}}_c$-$\mathbf{3}_c$ & 3028,~100\% & 858  & 398 & $64$   & $-926$    & $2$   & $7$   & $-15$  & $-61$ &0.79&0.85&0.71&1.07&0.63\\
\noalign{\smallskip}
\toprule[0.8pt]
\noalign{\smallskip}
  &   &                        & $\bar{\mathbf{3}}_c$-$\mathbf{3}_c$ & 3013,~94\%  & 1294 & 405 & $-191$ & $-967$  & $-25$  & $0$   & $-136$  & $-68$ &0.76&0.69&0.86&1.15&0.84\\
  & 0 & $01^{-}$               & $\mathbf{6}_c$-$\bar{\mathbf{6}}_c$ & 3270,~6\%   & 1027 & 469 & $-123$ & $-752$  & $2$    & $-11$ & $7$     & $-50$ &0.83&0.93&0.79&1.16&0.70\\
  &   &                        & C.C.                                & 2992        & 1316 & 393 & $-229$ & $-970$  & $-23$  & $-1$  & $-126$  & $-69$ &0.75&0.70&0.86&1.13&0.81\\
\noalign{\smallskip}
  &   &                        & $\bar{\mathbf{3}}_c$-$\mathbf{3}_c$ & 3055,~99\%  & 1251 & 421 & $-147$ & $-943$  & $-25$  & $0$   & $-136$  & $-67$ &0.80&0.69&0.89&1.17&0.85\\
1 & 1 & $00^{-},01^{-},02^{-}$ & $\mathbf{6}_c$-$\bar{\mathbf{6}}_c$ & 3393,~1\%   & 901  & 524 & $1$    & $-697$  & $1$    & $0$   & $6$     & $-44$ &0.89&0.96&0.84&1.23&0.75\\
  &   &                        & C.C.                                & 3051        & 1255 & 418 & $-154$ & $-944$  & $-24$  & $0$   & $-134$  & $-67$ &0.80&0.69&0.88&1.17&0.84\\
\noalign{\smallskip}
  & 2 & $01^{-},02^{-},03^{-}$ & $\mathbf{6}_c$-$\bar{\mathbf{6}}_c$ & 3432,~100\% & 865  & 541 & $33$   & $-677$  & $1$    & $4$   & $6$     & $-42$ &0.90&0.97&0.86&1.25&0.77\\
\noalign{\smallskip}
\toprule[0.8pt]
\noalign{\smallskip}
  &   &                        & $\bar{\mathbf{3}}_c$-$\mathbf{3}_c$ & 3228,~92\%  & 970  & 470 & $-1$   & $-839$  & $3$    & $-9$  & $-15$  & $-51$ &0.79&0.86&0.89&1.20&0.83\\
  & 0 & $11^{-}$               & $\mathbf{6}_c$-$\bar{\mathbf{6}}_c$ & 3429,~8\%   & 878  & 544 & $28$   & $-691$  & $-2$   & $0$   & $11$   & $-41$ &0.90&1.00&0.85&1.25&0.76\\
  &   &                        & C.C.                                & 3208        & 1001 & 459 & $-40$  & $-843$  & $3$    & $-9$  & $-12$  & $-52$ &0.78&0.86&0.87&1.18&0.80\\
\noalign{\smallskip}
  &   &                        & $\bar{\mathbf{3}}_c$-$\mathbf{3}_c$ & 3231,~97\%  & 969  & 474 & $-9$   & $-843$  & $3$    & $0$  & $-14$  & $-50$ &0.77&0.87&0.92&1.21&0.85\\
1 & 1 & $10^{-},11^{-},12^{-}$ & $\mathbf{6}_c$-$\bar{\mathbf{6}}_c$ & 3416,~3\%   & 887  & 538 & $17$   & $-694$  & $-2$   & $0$  & $11$   & $-42$ &0.88&1.00&0.85&1.24&0.75\\
  &   &                        & C.C.                                & 3222        & 980  & 470 & $-23$  & $-844$  & $2$    & $0$  & $-13$  & $-51$ &0.77&0.87&0.91&1.20&0.87\\
\noalign{\smallskip}
  & 2 & $11^{-},12^{-},13^{-}$ & $\bar{\mathbf{3}}_c$-$\mathbf{3}_c$ & 3291,~100\% & 908  & 500 & $50$   & $-812$  & $2$    & $4$  & $-14$  & $-48$ &0.81&0.87&0.94&1.25&0.87\\                                                                                          \noalign{\smallskip}
\toprule[0.8pt]
\end{tabular}
\end{table*}

\subsection{$[cu][\bar{s}\bar{d}]$ spectrum and $T^a_{c\bar{s}}(2900)$}

Assuming the states $T^a_{c\bar{s}}(2900)^0$ and $T^a_{c\bar{s}}(2900)^{++}$
to belong to the same isospin triplet, we also investigate the properties of
the ground and $P$-wave states $[cu][\bar{s}\bar{d}]$ with various spin,
$U$-spin and color combinations in the MCFTM. Similar to the isospin of
the antidiquark $[\bar{u}\bar{d}]$ in the state $[cs][\bar{u}\bar{d}]$,
we consider the $U$-spin of the antidiquark $[\bar{s}\bar{d}]$ in the
state $[cu][\bar{s}\bar{d}]$. In this way, we can define $U=0$ for the
$U$-spin antisymmetric $[\bar{s}\bar{d}]$ and $U=1$ for the $U$-spin
symmetric $[\bar{s}\bar{d}]$. In the same way, we can also define the $V$-spin
for the state $[cd][\bar{s}\bar{u}]$. Numerical results for the states
$[cu][\bar{s}\bar{d}]$ are presented in Table~\ref{cusd}. It can be found
from Tables~\ref{csud} and \ref{cusd} that the states $[cs][\bar{u}\bar{d}]$
and $[cu][\bar{s}\bar{d}]$ have a similar spectrum.

In the low-spin $(S\leq1)$ states $[cu][\bar{s}\bar{d}]$ and
$[cs][\bar{u}\bar{d}]$, the magnitude of the $\pi$-meson and $K$-meson exchange
interactions are distinguished, which results in their mass difference.
The masses of the states $[cu][\bar{s}\bar{d}]$ with $U=1$ are slightly
lower than those of the states $[cs][\bar{u}\bar{d}]$ with $I=1$, which
mainly originates from the different contribution from the $K$-meson
exchange interaction. In the states $[cu][\bar{s}\bar{d}]$ with $U=1$,
the interaction can provide a small attraction while the interaction
vanishes in the states $[cs][\bar{u}\bar{d}]$. However, the states
$[cu][\bar{s}\bar{d}]$ with $U=0$ are much higher than those of the
states $[cs][\bar{u}\bar{d}]$ with $I=0$ because of the strong attraction
induced by the $\pi$-meson exchange interaction. The high-spin $(S=2)$
states $[cu][\bar{s}\bar{d}]$ and $[cs][\bar{u}\bar{d}]$ are almost
degenerate because both the $\pi$-meson and $K$-meson exchange interactions
are very weak.

Using the QCD sum rules, the doubly charged states $[sd][\bar{u}\bar{c}]$ with the
spin-parity of $0^+$, $0^-$ and $1^+$~ have been explored \cite{agaev2017}. The
states with $0^+$ and $1^+$ have masses of $2628^{+166}_{-153}$ MeV and $2826^{+134}_{-157}$
MeV~\cite{agaev2017}, respectively, which are consistent with the corresponding
results in the present work within the error range. The mass of the state
with $0^-$ is $2719^{+144}_{-156}$ MeV~\cite{agaev2017}, which is much lower
than that of the state in the present work. Using QCD sum rules, the state
$[sd][\bar{u}\bar{c}]$ with $1^-$ have been investigated and gave a mass of $3515\pm125$
MeV~\cite{agaev20213}, which is much higher than the model prediction on the
state.

The mass of the state $[cu]_{\bar{\mathbf{3}_c}}[\bar{s}\bar{d}]_{\mathbf{3}_c}$
with $U(J^P)=1(0^+)$ is 2923 MeV, see Table~\ref{cusd}, which is highly consistent
with those of the states $T^a_{c\bar{s}0}(2900)^{0}$ and $T^a_{c\bar{s}0}(2900)^{++}$
reported by the LHCb Collaboration. The state is a compact state composed of an
axial-vector $[cu]_{\bar{\mathbf{3}}_c}$ and an axial-vector $[\bar{s}\bar{d}]_{\mathbf{3}_c}$.
The state $[cu]_{\mathbf{6}_c}[\bar{s}\bar{d}]_{\bar{\mathbf{6}_c}}$ with $U(J^P)=1(0^+)$
is much higher than those of the states $T^a_{c\bar{s}0}(2900)^{0}$ and
$T^a_{c\bar{s}0}(2900)^{++}$. After coupling two color configurations, the
mass of the state $[cu][\bar{s}\bar{d}]$ with $U(J^P)=1(0^+)$ can be
decreased to 2837 MeV, which is slightly lighter than those of the states
$T^a_{c\bar{s}0}(2900)^{0}$ and $T^a_{c\bar{s}0}(2900)^{++}$. Therefore, the
state $[cu][\bar{s}\bar{d}]$ with $U(J^P)=1(0^+)$ may be the main component
of the states $T^a_{c\bar{s}0}(2900)^{0}$ and $T^a_{c\bar{s}0}(2900)^{++}$.
The state $[cu][\bar{s}\bar{d}]$ with $U(J^P)=0(0^+)$, the partner of the states
$T^a_{c\bar{s}0}(2900)^{0}$ and $T^a_{c\bar{s}0}(2900)^{++}$, may exist
and has a mass of about 2583 MeV in the model.

\section{summary}

With the Gaussian expansion method as a high precision method, in this work we employ
the multiquark color flux-tube model to perform a systematically investigation on
the properties of the ground and $P$-wave states $[cs][\bar{u}\bar{d}]$ and
$[cu][\bar{s}\bar{d}]$ with various spin, isospin or $U$-spin and color combinations
in the present work. The model includes a multibody confinement potential, the
one-gluon-exchange interaction, the one-boson-exchange interaction ($\pi$, $K$
and $\eta$), and the $\sigma$-meson exchange interaction. The multi-body confinement
potential is a collective degree of freedom, which can bind all particles
together to establish a compact state. The states $[cs][\bar{u}\bar{d}]$ and
$[cu][\bar{s}\bar{d}]$ have similar mass spectra in the model. The mass
difference between two states mainly originates from the different magnitudes
of the $\pi$-meson and $K$-meson exchange interactions in the states.

Matching our results with the spin-parity and mass of the states $T_{cs0}(2900)^0$
and $T_{cs1}(2900)^0$ reported by the LHCb Collaboration, we can describe them as
the compact states $[cs][\bar{u}\bar{d}]$ with $I(J^{P})=1(0^+)$ and $0(1^-)$
in the model, respectively. The ground state $T_{cs0}(2900)^0$ is mainly made
of strongly overlapped an axial-vector $[cs]_{\bar{\mathbf{3}}_c}$ and an axial-vector
$[\bar{u}\bar{d}]_{\mathbf{3}_c}$.  If the state $T_{cs0}(2900)^0$ really belongs
to an isotriplet within the diquark-antidiquark picture, its charged partners would
be abundant in the model. The $P$-wave state $T_{cs1}(2900)^0$ is dominantly
consisting of a gradually separated scalar or axial vector $[cs]_{\bar{\mathbf{3}}_c}$
and a scalar $[\bar{u}\bar{d}]_{\mathbf{3}_c}$ in the shape of a dumbbell. In
addition, the states $[cs][\bar{u}\bar{d}]$ with $I(J^P)=0(0^+)$ and $0(1^+)$ may
exist and the predicted masses are about 2500-2600 MeV.

The predicted mass of the state $\left [[cu]_{\bar{\mathbf{3}}_c}[\bar{s}\bar{d}]_
{\mathbf{3}_c}\right ]_{\mathbf{1}_c}$ with $U(J^P)=1(0^+)$ in the model is
in good agreement with that of the states $T^a_{c\bar{s}0}(2900)^{0}$ and
$T^a_{c\bar{s}0}(2900)^{++}$. After considering the coupling of two color
configurations, the state $[cu][\bar{s}\bar{d}]$ is slightly lighter than
the states $T^a_{c\bar{s}0}(2900)^{0}$ and $T^a_{c\bar{s}0}(2900)^{++}$.
In this way, we cannot exclude the possibility that the state $[cu][\bar{s}\bar{d}]$
with $U(J^P)=1(0^+)$ may be the main component of the states
$T^a_{c\bar{s}0}(2900)^{0}$ and $T^a_{c\bar{s}0}(2900)^{++}$ in the model.
The state $[cu][\bar{s}\bar{d}]$ with $U(J^P)=0(0^+)$, the partner of the
states $T^a_{c\bar{s}0}(2900)^{0}$ and $T^a_{c\bar{s}0}(2900)^{++}$, may exist
and has a predicted mass of about 2583 MeV.

Hopefully, the systematical investigation on the states $[cs][\bar{u}\bar{d}]$
and $[cu][\bar{s}\bar{d}]$ will be useful for the understanding of the properties
of the exotic states $T_{cs}(2900)$ and $T^a_{c\bar{s}}(2900)$ and the search of
the new tetraquark states. We also expect more experimental and theoretical
investigations to verify and understand the tetraquark states in the future.

\acknowledgments

{This work is partly supported by the Chongqing Natural Science Foundation under
Project No. cstc2021jcyj-msxmX0078, and Fundamental Research Funds for
the Central Universities under Contracts No. SWU118111.}


\begin{thebibliography}{99}
\bibitem{experement} R. Aaij et al. (LHCb Collaboration), Phys. Rev. Lett. \textbf{125}, 242001 (2020);
  R. Aaij et al. (LHCb Collaboration), Phys. Rev. D \textbf{102}, 112003 (2020).
\bibitem{rename} LHCb Collaboration, arXiv: 2206.15233 [hep-ex].
\bibitem{experement1} https://indico.cern.ch/event/1176505/.
\bibitem{chen2020} H.X. Chen, W. Chen, R.R. Dong, and N. Su, Chin. Phys. Lett. \textbf{37}, 101201 (2020).
\bibitem{molina2020} R. Molina and E. Oset, Phys. Lett. B \textbf{811}, 135870 (2020).
\bibitem{liu2020} M.Z. Liu, J.J. Xie, and L.S. Geng, Phys. Rev. D \textbf{102}, 091502(R)  (2020).
\bibitem{molina2021} M.W. Hu, X.Y. Lao, P. Ling, and Q. Wang, Chin. Phys. C \textbf{45}, 021003 (2021).
\bibitem{xiao2021} C.J. Xiao, D.Y. Chen, Y.B. Dong, and G.W. Meng, Phys. Rev. D \textbf{103}, 034004 (2021).
\bibitem{agaev2021} S.S. Agaev, K. Azizi, and H. Sundu, J. Phys. G \textbf{48}, 085012 (2021).
\bibitem{xue2021} Y.Y. Xue, X. Jin, H.X. Huang, J.L. Ping, and F. Wang, Phys. Rev. D \textbf{103}, 054010 (2021).
\bibitem{liu2022} B. Wang and S.L. Zhu, Eur. Phys. J. C \textbf{82}, 419 (2022).
\bibitem{dai2022} L.R. Dai, R. Molina, and E. Oset, Phys. Lett. B \textbf{832}, 137219 (2022);
  Phys. Rev. D \textbf{105}, 096022 (2022).
\bibitem{he2021} J. He and D.Y. Chen, Chin. Phys. C \textbf{45}, 063102 (2021).
\bibitem{qi2021} J.J. Qi, Z.Y. Wang, Z.F. Zhang, and X.H. Guo, Eur. Phys. J. C \textbf{81}, 639 (2021).
\bibitem{chenh2021} H. Chen, H.R. Qi, and H.Q. Zheng, Eur. Phys. J. C \textbf{81}, 812 (2021).
\bibitem{agaev2022} S.S. Agaev, K. Azizi, and H. Sundu, arXiv: 2207.02648 [hep-ph].
\bibitem{rchen2022} R. Chen and Q. Huang, arXiv: 2208.10196 [hep-ph].
\bibitem{agaev20212} S.S. Agaev, K. Azizi, and H. Sundu, Nucl. Phys. A \textbf{1011}, 122202 (2021).
\bibitem{agaev20213} S.S. Agaev, K. Azizi, and H. Sundu, Eur. Phys. J. C \textbf{820}, 136530 (2021).
\bibitem{ozdem2022} U. \"{O}zdem, K. Azizi, Eur. Phys. J. A \textbf{58}, 171 (2022).
\bibitem{karliner2020} M. Karliner and J. L. Rosner, Phys. Rev. D \textbf{102}, 094016 (2020).
\bibitem{cheng2020} J.B. Cheng, S.Y. Li, Y.R. Liu, Y.N. Liu, Z.G. Si, and T. Yao, Phys. Rev. D \textbf{101}, 114017 (2020).
\bibitem{zhang2021} J.R. Zhang, Phys. Rev. D \textbf{103}, 054019 (2021).
\bibitem{wang2021} G.J. Wang, L. Meng, L.Y. Xiao, M. Oka, and S.L. Zhu, Eur. Phys. J. C \textbf{81} 188 (2021).
\bibitem{lv2020} Q.F. L\"{u}, D.Y. Chen, and Y.B. Dong, Phys. Rev. D \textbf{102}, 074021 (2020).
\bibitem{wang2020} Z.G. Wang, Int. J. Mod. Phys. A \textbf{35}, 2050187 (2020).
\bibitem{yang2021} G. Yang, J. Ping, and J. Segovia, Phys. Rev. D \textbf{103}, 074011 (2021).
\bibitem{guo2022} T. Guo , J.N Li , J.X. Zhao, and L.Y. He, Phys. Rev. D \textbf{105}, 054018 (2022).
\bibitem{albuquerque2021} R.M. Albuquerque, S. Narison, D. Rabetiarivony, and G. Randriamanatrika, Nucl. Phys. A \textbf{1007}, 122113 (2021).
\bibitem{triangle} X.H. Liu, M.J. Yan, H.W. Ke, G. Li, and J.J. Xie, Eur. Phys. J. C \textbf{80}, 1178 (2020).
\bibitem{triangle1} Y.H. Ge, X.H. Liu, and H.W. Ke, Eur. Phys. J. C \textbf{82}, 955 (2022).
\bibitem{burns} T.J. Burns and E.S. Swanson, Phys. Lett. B \textbf{813}, 136057 (2021).
\bibitem{chen2022} H.X. Chen, W. Chen, X. Liu, Y.R. Liu, and S.L. Zhu, arXiv: 2204.02649 [hep-ph].
\bibitem{deng2020} C.R. Deng, H. Chen, and J.L. Ping, Phys. Rev. D \textbf{101}, 054039 (2020).
\bibitem{latt1} T.T. Takahashi, H. Suganuma, Y. Nemoto, and H. Matsufuru, Phys. Rev. D \textbf{65}, 114509 (2002).
\bibitem{latt2} F. Okiharu, H. Suganuma, and T.T. Takahashi, Phys. Rev. Lett. \textbf{94}, 192001 (2005).
\bibitem{chiral} J. Vijande, F. Fernandez, and A. Valcarce, J. Phys. G \textbf{31}, 481 (2005).
\bibitem{kappa1} G.S. Bali, Phys. Rev. D \textbf{62}, 114503 (2000).
\bibitem{kappa2} C. Semay, Eur. Phys. J. A \textbf{22}, 353 (2004).
\bibitem{kappa3} N. Cardoso, M. Cardoso, and P. Bicudo, Phys. Lett. B \textbf{710}, 343 (2012).
\bibitem{chnqm} A. Manohar and H. Georgi, Nucl. Phys. B \textbf{234}, 189 (1984).
\bibitem{masssigma} M.D. Scadron, Phys. Rev. D \textbf{26}, 239 (1982).
\bibitem{GEM} E. Hiyama, Y. Kino, and M. Kamimura, Prog. Part. Nucl. Phys. \textbf{51}, 223 (2003).
\bibitem{minuit} F. James and M. Roos, Comp. Phys. Comm. \textbf{10}, 343 (1975).
\bibitem{deng2020x} C.R. Deng, H. Chen, and J.L. Ping, Eur. Phys. J. A \textbf{56}, 9 (2020).
\bibitem{diquark} C.R. Deng and S.L. Zhu, Sci. Bull. \textbf{67}, 1522 (2022).
\bibitem{deng2021} C.R. Deng, H. Chen, and J.L. Ping, Phys. Rev. D \textbf{103}, 014001 (2021).
\bibitem{agaev2017} S.S. Agaev, K. Azizi, and H. Sundu, Phys. Lett. B \textbf{78}, 141 (2018).


\end{thebibliography}
\end{document}